\documentclass[10pt,journal,compsoc]{IEEEtran}

\ifCLASSOPTIONcompsoc
  \usepackage[nocompress]{cite}
\else
  \usepackage{cite}
\fi

\usepackage[pdftex]{graphicx}
\usepackage{amsmath}
\usepackage{algorithmic}
\usepackage{array}
\usepackage{stfloats}
\usepackage[hyphens]{url}
\usepackage{glossaries}
\usepackage{listings}
\usepackage{xspace}
\usepackage{interval}
\usepackage{orcidlink}

\begin{document}

\title{Taming Offload Overheads in a Massively\\Parallel Open-Source RISC-V MPSoC:\\Analysis and Optimization}

\author{Luca~Colagrande\orcidlink{0000-0002-7986-1975},~\IEEEmembership{Graduate Student Member,~IEEE,}
Luca~Benini\orcidlink{0000-0001-8068-3806},~\IEEEmembership{Fellow,~IEEE,}
\IEEEcompsocitemizethanks{\IEEEcompsocthanksitem L. Colagrande and L. Benini are with the Integrated
Systems Laboratory (IIS), Swiss Federal Institute of Technology, Zurich, Switzerland.\protect\\
E-mail: \{colluca,lbenini\}@iis.ee.ethz.ch
\IEEEcompsocthanksitem L. Benini is also with the Department of Electrical, Electronic and Information
Engineering (DEI), University of Bologna, Bologna, Italy}
}


\IEEEtitleabstractindextext{%
\begin{abstract}
Heterogeneous multi-core architectures combine on a single chip a few large, general-purpose \textit{host} cores, optimized for single-thread performance, with (many) clusters of small, specialized, energy-efficient \textit{accelerator} cores for data-parallel processing.
Offloading a computation to the many-core acceleration fabric implies synchronization and communication overheads which can hamper overall performance and efficiency, particularly for small and fine-grained parallel tasks.
In this work, we present a detailed, cycle-accurate quantitative analysis of the offload overheads on Occamy, an open-source massively parallel RISC-V based heterogeneous MPSoC.
We study how the overheads scale with the number of accelerator cores.
We explore an approach to drastically reduce these overheads by co-designing the hardware and the offload routines.
Notably, we demonstrate that by incorporating multicast capabilities into the Network-on-Chip of a large (200+ cores) accelerator fabric we can improve offloaded application runtimes by as much as 2.3x, restoring more than 70\% of the ideally attainable speedups.
Finally, we propose a quantitative model to estimate the runtime of selected applications accounting for the offload overheads, with an error consistently below 15\%.
\end{abstract}

}

\maketitle

\newcommand{\ResultMcastOverheadMean}{185\xspace}
\newcommand{\ResultMcastOverheadStddev}{18\xspace}
\newcommand{\ResultOverheadSingleClusterMean}{242\xspace}
\newcommand{\ResultOverheadSingleClusterStddev}{65\xspace}
\newcommand{\ResultOverheadMatmulMax}{1146\xspace}
\newcommand{\ResultOverheadThirtyTwoClusterStddev}{256\xspace}
\newcommand{\ResultIdealSpeedupMatmulMax}{3.02\xspace}
\newcommand{\ResultIdealSpeedupBFSMax}{1.31\xspace}
\newcommand{\ResultIdealSpeedupFractionAXPYMonteCarloMatmulMin}{70\xspace}
\newcommand{\ResultIdealSpeedupFractionAXPYMonteCarloMatmulMax}{90\xspace}
\newcommand{\ResultIdealSpeedupFractionATAXCovarianceBFSMin}{85\xspace}
\newcommand{\ResultIdealSpeedupFractionATAXCovarianceBFSMax}{96\xspace}
\newcommand{\ResultMaxSpeedup}{2.3\xspace}
\newcommand{\ResultMaxModelingErrorRounded}{15\xspace}


\newacronym[plural=WANs, firstplural={Wide Area Networks (WANs)}]{wan}{WAN}{Wide Area Network}
\newacronym[plural=WSNs, firstplural={Wireless Sensor Networks (WSNs)}]{wsn}{WSN}{Wireless Sensor Network}
\newacronym{simd}{SIMD}{Single Instruction Multiple Data}
\newacronym{os}{OS}{Operating System}
\newacronym{ble}{BLE}{Bluetooth Low-Energy}
\newacronym{wifi}{Wi-FI}{Wireless Fidelity}
\newacronym[plural=DVS, firstplural={Dynamic Vision Sensors (DVS)}]{dvs}{DVS}{Dynamic Vision Sensor}
\newacronym{ptz}{PTZ}{Pan-Tilt Unit}

\newacronym[plural=FLLs,firstplural=Frequency Locked Loops (FLLs)]{fll}{FLL}{Frequency Locked Loop}
\newacronym{dram}{DRAM}{Dynamic Random Access Memory}
\newacronym{fpu}{FPU}{Floating Point Unit}
\newacronym{dma}{DMA}{Direct Memory Access}
\newacronym[plural=LUTs, firstplural={Lookup Tables (LUTs)}]{lut}{LUT}{Lookup Table}
\newacronym[plural=FPGAs, firstplural={Field Programmable Gate Arrays (FPGAs)}]{fpga}{FPGA}{Field Programmable Gate Array}
\newacronym{dsp}{DSP}{Digital Signal Processing}
\newacronym{mcu}{MCU}{Microcontroller Unit}
\newacronym{spi}{SPI}{Serial Peripheral Interface}
\newacronym{cpi}{CPI}{Camera Parallel Interface}
\newacronym{fifo}{FIFO}{First-In First-Out Queue}
\newacronym{uart}{UART}{Universal Asynchronous Receiver-Transmitter}
\newacronym{raw}{RAW}{Read-After-Write}
\newacronym[plural=ISAs, firstplural={Instruction Set Architectures (ISAs)}]{isa}{ISA}{Instruction Set Architecture}
\newacronym{xbar}{XBAR}{crossbar}
\newacronym[firstplural=Scratch-Pad Memories (SPMs)]{spm}{SPM}{Scratch-Pad Memory}
\newacronym{ppa}{PPA}{Power Performance Area}
\newacronym{ipi}{IPI}{Inter-Processor Interrupt}
\newacronym[firstplural=Software-Generated Interrupts (SGIs)]{sgi}{SGI}{Software-Generated Interrupt}
\newacronym{pe}{PE}{Processing Element}
\newacronym{tcdm}{TCDM}{Tightly-Coupled Data Memory}
\newacronym{lsu}{LSU}{Load-Store Unit}
\newacronym{icache}{I\$}{Instruction Cache}
\newacronym{dcache}{D\$}{Data Cache}
\newacronym{wfi}{WFI}{Wait For Interrupt}

\newacronym{ste}{STE}{Straight-Through-Estimator}

\newacronym[plural=PTUs, firstplural={Pan-Tilt Units}]{ptu}{PTU}{Pan-Tilt Unit}
\newacronym{mdf}{MDF}{Medium-density fibreboard}
\newacronym{cvat}{CVAT}{Computer Vision Annotation Tool}
\newacronym{coco}{COCO}{Common Objects in Context}
\newacronym{soa}{SoA}{State of the Art}
\newacronym{sf}{SF}{Sensor Fusion}

\newacronym{dl}{DL}{Deep Learning}
\newacronym{bn}{BN}{Batch Normalization}
\newacronym{FGSM}{FBK}{Fast Gradient Sign Method}
\newacronym{lr}{LR}{Learning Rate}
\newacronym{sgd}{SGD}{Stochastic Gradient Descent}
\newacronym{gd}{GD}{Gradient Descent}

\newacronym{sta}{STA}{Static Timing Analysis}

\newacronym[plural=GPIOs, firstplural={General Purpose Inupt Outputs (GPIOs)}]{gpio}{GPIO}{General Purpose Input Output}
\newacronym[plural=LDOs, firstplural={Low Dropout Regulators (LDOs)}]{ldo}{LDO}{Low Dropout Regulator}

\newacronym{inq}{INQ}{Incremental Network Quantization}

\newacronym{CV}{CV}{Computer Vision}
\newacronym{EoT}{EoT}{Expectation over Transformation}
\newacronym{RPN}{RPN}{Region Proposal Network}
\newacronym{TV}{TV}{Total Variation}
\newacronym{NPS}{NPS}{Non-Printability Score}
\newacronym{STN}{STN}{Spatial Transformer Network}
\newacronym{MTCNN}{MTCNN}{Multi-Task Convolutional Neural Network}
\newacronym{YOLO}{YOLO}{You Only Look Once}
\newacronym{SSD}{SSD}{Single Shot Detector}
\newacronym{SOTA}{SOTA}{State of the Art}
\newacronym{NMS}{NMS}{Non-Maximum Suppression}
\newacronym{ic}{IC}{Integrated Circuit}
\newacronym{rf}{RF}{Radio Frequency}
\newacronym{tcxo}{TCXO}{Temperature Controlled Crystal Oscillator}
\newacronym{jtag}{JTAG}{Joint Test Action Group industry standard}
\newacronym{swd}{SWD}{Serial Wire Debug}
\newacronym{sdio}{SDIO}{Serial Data Input Output}

\newacronym[plural=PCBs, firstplural={Printed Circuit Boards (PCB)}]{pcb}{PCB}{Printed Circuit Board}
\newacronym[plural=ASICs, firstplural={Application Specific Integrated Circuits}]{asic}{ASIC}{Application Specific Integrated Circuit}

\newacronym[plural=BNNs, firstplural={Binary Neural Networks (BNNs)}]{bnn}{BNN}{Binary Neural Network}
\newacronym[plural=NNs, firstplural={Neural Networks}]{nn}{NN}{Neural Network (NNs)}
\newacronym[plural=SCMs, firstplural={Standard Cell Memories (SCMs)}]{scm}{SCM}{Standard Cell Memory}
\newacronym{ann}{ANN}{Artificial Neural Networks}
\newacronym{ml}{ML}{Machine Learning}
\newacronym{ai}{AI}{Artificial Intelligence}
\newacronym{iot}{IoT}{Internet of Things}
\newacronym{fft}{FFT}{Fast Fourier Transform}
\newacronym[plural=OCUs, firstplural={Output Channel Compute Units (OCUs)}]{ocu}{OCU}{Output Channel Compute Unit}
\newacronym{alu}{ALU}{Arithmetic Logic Unit}
\newacronym{mac}{MAC}{Multiply-Accumulate}
\newacronym[firstplural={systems-on-chip (SoCs)}]{soc}{SoC}{system-on-chip}
\newacronym[firstplural={multi-processor systems-on-chip (MPSoCs)}]{mpsoc}{MPSoC}{multi-processor system-on-chip}

\newacronym{PGD}{PGD}{Projected Gradient Descend}
\newacronym{CW}{CW}{Carlini-Wagner}
\newacronym{OD}{OD}{Object Detection}

\newacronym{rrf}{RRF}{RADAR Repetition Frequency}
\newacronym{nlp}{NLP}{Natural Language Processing}
\newacronym{qam}{QAM}{Quadrature Amplitude Modulation}
\newacronym{rri}{RRI}{RADAR Repetition Interval}
\newacronym{radar}{RADAR}{Radio Detection and Ranging}
\newacronym{loocv}{LOOCV}{Leave-one-out cross validation}

\newacronym{bsp}{BSP}{Board Support Package}
\newacronym{ttn}{TTN}{The Things Network}
\newacronym{wip}{WIP}{Work in Progress}
\newacronym{json}{JSON}{JavaScript Object Notation}
\newacronym{qat}{QAT}{Quantization-Aware Training}

\newacronym{cls}{CLS}{Classification Error}
\newacronym{loc}{LOC}{Localization Error}
\newacronym{bkgd}{BKGD}{Background Error}
\newacronym{roc}{ROC}{Receiver Operating Characteristic}
\newacronym{frr}{FRR}{False Rejection Rate}
\newacronym{eer}{EER}{Equal Error Rate}
\newacronym{snr}{SNR}{Signal-to-Noise Ratio}
\newacronym{flop}{FLOP}{Floating-Point Operation}
\newacronym{fp}{FP}{Floating-Point}
\newacronym{fps}{FPS}{Frames Per Second}

\newacronym{gsc}{GSC}{Google Speech Commands}
\newacronym{mswc}{MSWC}{Multilingual Spoken Words Corpus}
\newacronym{demand}{DEMAND}{Diverse Environments Multichannel Acoustic Noise Database}

\newacronym[plural=SNNs, firstplural={Spiking Neural Networks (SNNs)}]{snn}{SNN}{Spiking Neural Network}
\newacronym[plural=DNNs, firstplural={Deep Neural Networks (DNNs)}]{dnn}{DNN}{Deep Neural Network}
\newacronym[plural=TCNs,firstplural=Temporal Convolutional Networks]{tcn}{TCN}{Temporal Convolutional Network}
\newacronym[plural=CNNs,firstplural=Convolutional Neural Networks (CNNs)]{cnn}{CNN}{Convolutional Neural Network}
\newacronym[plural=TNNs,firstplural=Ternarized Neural Networks]{tnn}{TNN}{Ternarized Neural Network}
\newacronym{ds-cnn}{DS-CNN}{Depthwise Separable Convolutional Neural Network}
\newacronym{rnn}{RNN}{Recurrent Neural Network}
\newacronym{gcn}{GCN}{Graph Convolutional Network}
\newacronym{mhsa}{MHSA}{Multi-Head Self Attention}
\newacronym{crnn}{CRNN}{Convolutional Recurrent Neural Network}
\newacronym{clca}{CLCA}{Convolutional Linear Cross-Attention}

\newacronym{bf}{BF}{Beamforming}
\newacronym{anc}{ANC}{Active Noise Cancellation}
\newacronym{agc}{AGC}{Automatic Gain Control}
\newacronym{se}{SE}{Speech Enhancement}
\newacronym{mct}{MCT}{Multi-Condition Training}
\newacronym{mcta}{MCTA}{Multi-Condition Training \& Adaptation}
\newacronym{pcen}{PCEN}{Per-Channel Energy Normalization}
\newacronym{mfcc}{MFCC}{Mel-Frequency Cepstral Coefficient}
\newacronym{asr}{ASR}{Automated Speech Recognition}
\newacronym{kws}{KWS}{Keyword Spotting}
\newacronym{odl}{ODL}{On-Device Learning}


\newacronym{nl-kws}{NL-KWS}{Noiseless Keyword Spotting}
\newacronym{na-kws}{NA-KWS}{Noise-Aware Keyword Spotting}
\newacronym{odda}{ODDA}{On-Device Domain Adaptation}
\newacronym{hpm}{HPM}{High-Performance Mode}
\newacronym{lpm}{LPM}{Low-Power Mode}

\lstset{
language=C,                        
basicstyle=\ttfamily\footnotesize, 
literate={~}{{\fontfamily{ptm}\selectfont \textasciitilde}}1
}
\bstctlcite{IEEEexample:BSTcontrol}

\IEEEdisplaynontitleabstractindextext
\IEEEpeerreviewmaketitle
\IEEEraisesectionheading{\section{Introduction}\label{sec:introduction}}

\IEEEPARstart{W}{ith} the end of Dennard scaling, computer architects have been increasingly relying on system specialization techniques to tackle computing's energy bottlenecks \cite{horowitz2014} and drive system performance forward \cite{taylor2012}. Heterogeneous computing is one way to exploit specialization, by integrating different compute units tailored to a diverse range of applications onto a single system \cite{schulte2015,zahran2019}. The specialization degree of the compute units can vary from dedicated accelerators to customized instruction processing cores of varying complexity \cite{biglittle}, and is subject to a trade-off between efficiency and flexibility \cite{nowatzki2016, fuchs2019}. In this work, we focus on a very common architectural pattern: heterogeneous (or asymmetric) multi-core architectures \cite{kumar2005}.
Specifically, we focus on \glspl{mpsoc} which couple one or more large general-purpose OS-capable cores (the \textit{host}) with many, small energy-efficient cores specialized for data-parallel workloads \cite{occamy, rossi2017} (the \textit{accelerator} or \textit{device}). These systems enable the execution of intensive data-parallel computations at high energy-efficiency and performance on the accelerator. At the same time, they achieve high single-thread performance for non-parallelizable code on the host.

Most heterogeneous system programming languages and frameworks assume the following platform model: the host is managing the computation and hands off suitable parts of it, designated by the programmer, to the accelerators. CUDA, OpenCL, OpenACC and OpenMP, since version 4.0, all adopt this execution model for heterogeneous programming.


In this context, we refer to the process of handing over parts of the computation to the accelerator as \textit{job offloading}.
It is the programmer's responsibility to define the workload partition between the host and the accelerator(s), i.e. to define jobs to offload to the accelerator(s). Making an optimal offload decision is non-intuitive \cite{che2009} and complicated by the fact that offloading introduces several overheads to the execution time and energy consumption of the job on the accelerator which must be taken into consideration. To this end, \textit{understanding ``when'', ``where'' and ``how'' these overheads form and being able to quantitatively estimate them is key}.

The offload decision is not only about determining whether a portion of the workload can benefit or not from offloading. In the case of GPU offloading for instance, the programmer has to specify also other parameters, such as the number of threads per block, which may have a significant influence on performance \cite{araujo2023}. In this case, the offload decision is non-binary; in addition to establishing \textit{if} a job is suitable for offload, the question \textit{how} to offload the job has to be answered as well. In the context of heterogeneous \glspl{mpsoc}, the number of cores to employ for a job represents such a parameter.



In fact, codes with a small amount of parallelism will not benefit from executing on all accelerator cores. We refer to this degree of parallelism as the \textit{width} of a parallel algorithm \cite{feo1988}. In order to enable efficient execution of both \textit{wide} and \textit{narrow} parallel codes, it is desirable to have a flexible offload mechanism which can selectively activate a subset of accelerator cores based on the job requirements. At the same time, parallel programs can be characterized by their \textit{length}\cite{feo1988} or \textit{granularity} \cite{chen1990}. Offloading too short or fine-grained programs might prove inefficient, if the offload overheads exceed the speedup or energy-efficiency gains from parallelization. In many cases, the width and length of a program are inversely related \cite{chen1990}. As we will see in section \ref{sec:breakdown}, part of the offload overheads in heterogeneous \glspl{mpsoc} grow with the number of cores selected for offload. Mapping a program onto a smaller number of cores might thus prove effective to reduce the overheads and justify offloading, however as a result the length of the program increases. Inevitably, this approach trades off performance by taking advantage of a reduced amount of parallelism in the computation. Thus, \textit{to exploit the full potential of heterogeneous architectures, that is to enable fine-grained heterogeneous execution, offload overheads must be reduced to a minimum}.


To favour a deep insight on hardware and software, eliminating error-prone and approximate reverse engineering efforts on proprietary hardware, and to provide a platform for future research and reproducible benchmarking, we develop our contributions on an open-source state-of-the-art RISC-V based heterogeneous \gls{mpsoc}. Summing up the main contributions of this paper, we:
\begin{enumerate}
\item Develop a fully open-source heavily-optimized bare-metal software framework implementing a host-centric heterogeneous model of execution on a state-of-the-art RISC-V based heterogeneous \gls{mpsoc}, analyzing both hardware and software. We use this framework to evaluate how the offload overheads affect the speedup of offloaded kernels and demonstrate the importance of reducing the offload overheads and making correct offload decisions.
\item Propose hardware extensions, co-designed with the software, to reduce the offload overheads. We evaluate how, by means of these extensions, we can restore part of the ideally attainable speedups on the accelerator.
\item Present a detailed cycle-accurate quantitative breakdown of the offload overheads in a heterogeneous \gls{mpsoc}. To the best of our knowledge, no previous work in literature conducts such an analysis, so our work aims to set a transparent and fully open-source
baseline for future comparisons. \footnote{https://github.com/colluca/occamy/tree/offload}
\item Based on our measurements, we develop an accurate and explainable analytical model to estimate the runtime of offloaded applications, and empirically validate the accuracy of our model on a variety of problem sizes. This model could be used to formulate the offload decision as an optimization problem and analytically derive optimal offload parameters.
\end{enumerate}

This work is an extension of a prior work-in-progress exploratory paper \cite{colagrande2024}.
Most of the present content is original, providing comprehensive, in-depth presentations and illustrating new concepts which could not be covered in the scope of a work-in-progress paper.
A versatile software framework has been developed to support offloading arbitrary kernels, and the whole work rebased on Occamy \cite{occamy}, an improved and silicon-proven architecture derived from Manticore \cite{manticore}.
All of the contributions listed above are entirely novel or present new insights supported by newly conducted experiments.

In this section we introduced the context, motivation and contributions of our work.
The remainder of this paper is structured as follows. Section \ref{sec:related} presents a review of state-of-the-art single-chip heterogeneous architectures, inter-processor interrupts and barrier synchronization infrastructure and related work on offload speedup modeling. In section \ref{sec:background} we describe the Occamy \gls{soc} architecture, as relevant for our discussion, and give a general overview of the offloading process. Section \ref{sec:implementation} describes the implementation of our offload routines and hardware extensions. In section \ref{sec:results} we present the results of this study, summarized in the contributions above.

\section{Related work}
\label{sec:related}

\subsection{Single-chip heterogeneous systems}

CPU-GPU solutions arguably represent the most common type of heterogeneous system, as can be seen from the large fraction of TOP500 and Green500 supercomputers adopting this architecture \cite{top500, green500}.
To tame the overhead incurred in transfering data off-chip, many recent processors integrate CPUs and GPUs on the same die (SoC) or multi-chiplet package (SiP) \cite{arora2020, ditty2014, gomes2022}.
As recent research shows, heterogeneous integrated \glspl{soc} lead to tangible performance and energy efficiency improvements on a variety of workloads \cite{hetherington2012, spafford2012, daga2011, lee2013}.
In parallel with this trend, many recent high-performance \glspl{soc} are adopting the RISC-V open instruction-set architecture (ISA) for their accelerator cores \cite{ignjatovic2022, ditzel2022, occamy, tine2021}, which makes for a sustainable development ecosystem and a solid base for wide-range adoption \cite{tine2021}. A large fraction of these accelerators adopt alternative execution models to the Single Instruction Multiple-Thread (SIMT) model embraced by GPUs \cite{ignjatovic2022, ditzel2022, occamy}, with the goal of further improving energy efficiency by eliminating thread scheduling hardware and costly context switching activity \cite{zaruba2021snitch}.
We develop this study on the Occamy \gls{soc} \cite{occamy}, as it combines the benefits of recent architectural trends with a fully open-source hardware design. This characteristic was fundamental to develop a cycle-accurate analysis and model of offload performance, whereas significant reverse-engineering efforts would otherwise be required, as evidenced by a multitude of previous studies \cite{khairy2020, wong2010, mei2017, fang2018, mei2014, stigt2022, jia2018, jia2019, zhang2017}. 


\subsection{Offload speedup modeling}

Offloading from host to accelerator is desirable when this improves some objective, may it be runtime, energy consumption or others. Modeling the objective as a function of kernel and offload parameters allows for making informed offload decisions and to select provably optimal offload parameters. Arguably the most studied objective is minimizing runtime or, equivalently, maximizing speedup.

Speedup models date as far back as Amdahl's law, which states that the speedup from parallelizing an application on multiple cores is limited by the serial fraction of the application. Amdahl's law provides a simple speedup model, applicable for workloads which can be decomposed into a serial fraction of constant runtime and a perfectly parallelizable fraction. Many works in literature have focused on extending the applicability of Amdahl's model to a wider range of applications \cite{li1988, sun1990, yavits2014} or architectures \cite{hill2008, rafiev2018}. However, very few works focus on the specific problem of modeling the speedup of an application offloaded onto an accelerator, accounting for offload overheads.

Pei et al. \cite{pei2016} propose speedup models for homogeneous asymmetric multicores and heterogeneous CPU-GPU multicores, among other architectures, taking offload (or ``data preparation'', in their terms) overheads into account. Their work is however purely theoretical. They extract application parameters (e.g. percentage of sequential load/store operations) to plug into their models from realistic heterogeneous applications taken from the Rodinia benchmark suite, but how useful and representative these metrics are is not demonstrated empirically.

The same authors published a follow-up to the previous paper \cite{pei2016-2}. To complement the previous efforts, they develop several microbenchmarks to measure the offload overheads. They perform some experiments to motivate the need of speedup models which account for these overheads. Specifically, they run seven kernels from the Rodinia benchmark suite on an Nvidia GTX 750 GPU. They measure the percentage of total execution time spent on kernel execution, CPU execution, CPU-GPU communication and synchronization. However, they do not build and validate any analytical model based on experimental data.

Similar measurements were also presented by the original authors of the Rodinia suite \cite{che2009}. They report the percentage of kernel execution, CPU execution, CPU-GPU communication and initial setup and I/O for each of the Rodinia kernels as measured on an Nvidia GTX 280. They do not report absolute runtime numbers, nor explain how the overheads form in the hardware and how to quantitatively model them.

Kannan et al. \cite{kannan2015} show that the performance of an application offloaded to the GPU largely depends on its arithmetic intensity and problem size (``working set size'', in their terms), and explain how this dependency relates to the offload overhead. They then propose a simple regression model to estimate the speedup on the GPU. As for the previous papers, they do not validate the accuracy of their model.

All these papers focus on discrete CPU-GPU heterogeneous architectures, where CPU and GPU reside on separate chips interconnected by a PCIe bus. Understandably, due to the proprietary closed-source nature of these architectures, it is difficult to evaluate the magnitude of the offload overheads as a whole and to model the offload runtime and speedup based on these measurements. In fact, all prior works fail to provide a precise quantitative breakdown of the offload overheads and to validate the accuracy of their models on experimental data.

With this work, we attempt to fill this gap by providing insight on the magnitude of the offload overheads on a real architecture and applications, explaining how these overheads originate in the software and hardware and developing an accurate runtime model based on this analysis.



\subsection{Inter-processor interrupts and barrier synchronization}

\begin{figure}[t]
\centering
\includegraphics[width=\columnwidth]{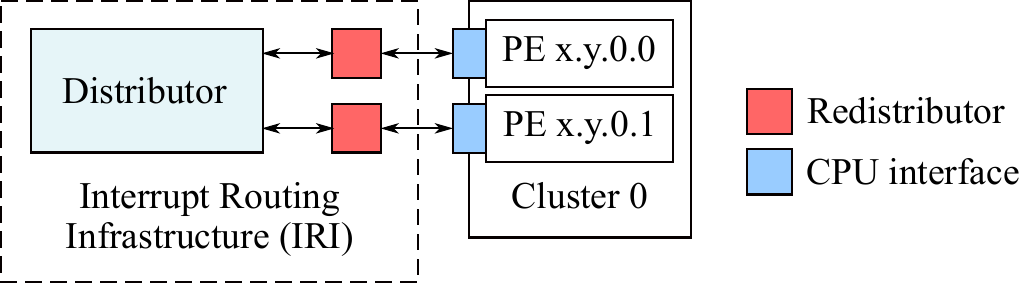}
\caption{Arm Generic Interrupt Controller (GIC) Architecture}
\label{fig:gic}
\end{figure}

As we will show, part of the offload overheads in a heterogeneous \gls{mpsoc} arise from \gls{ipi} delivery and barrier synchronization routines. We thus present a brief review of state-of-the-art \gls{ipi} and barrier infrastructure in comparison with our reference system, Occamy.

Unfortunately, much of the information on proprietary state-of-the-art systems is not disclosed. However, most ARM-based systems, such as the latest Fujitsu A64FX CPU designed for the Fugaku supercomputer \cite{yamamura2022}, comply with the ARM Generic Interrupt Controller (GIC) Architecture Specification \cite{arm_gic}. A block diagram of the GIC architecture is shown in figure \ref{fig:gic}. The specification mandates that \textit{\glspl{sgi}} initiated by a source \gls{pe} are routed to the target \glspl{pe} through a centralized \textit{Distributor}. This implies that \glspl{ipi} must travel the whole on-chip distance from the source \gls{pe} to a centralized distributor to the target \glspl{pe}, and cannot take shortcuts from one \gls{pe} to another. Furthermore, a \textit{CPU interface} for each \gls{pe} in the system provides the programming interface for acknowledging and deactivating interrupts and a corresponding \textit{Redistributor} for each \gls{pe} controls the pending and active states of \glspl{sgi}. From this, we assume that interrupts are cleared locally to every core without having to travel long on-chip distances to the centralized distributor. The GIC specification further provides facilities to broadcast \glspl{ipi} to all \glspl{pe} in the system or multicast to a maximum of 16 \glspl{pe} in the same \textit{Affinity level} 0 cluster.

Occamy's interrupt infrastructure is on par or improves upon the performance implications of the GIC specification.
A centralized interrupt controller (CLINT) is implemented in the peripherals domain of the system.
\glspl{ipi} can be triggered by any RISC-V hart in the system by writing to the memory-mapped MSIP (Machine Software Interrupt Pending) register in the CLINT.
To provide locally-clearable interrupts, Occamy implements a custom MCIP (Machine Cluster Interrupt Pending) register in every cluster.
These registers are mapped on the narrow network for access by any hart.
Access latency depends on the vicinity of the request initiator to the register, but is in any case lower than the latency to go through the centralized CLINT.
Cores have local low-latency access to their own MCIP bit.
These bits being packed in the same register for all cores in a cluster allows multicasting interrupts to all cores in a cluster, i.e. the equivalent of \glspl{pe} in the same affinity level 0 for the GIC specification, with a single store instruction.
Broadcasting \glspl{ipi} is not possible in Occamy.
However, in section \ref{sec:multicast} we describe an extension which allows versatile multicasting to all \glspl{pe} in the accelerator, representing a superset of the broadcast functionality.

\begin{figure*}[!t]
\centering
\includegraphics[width=\textwidth]{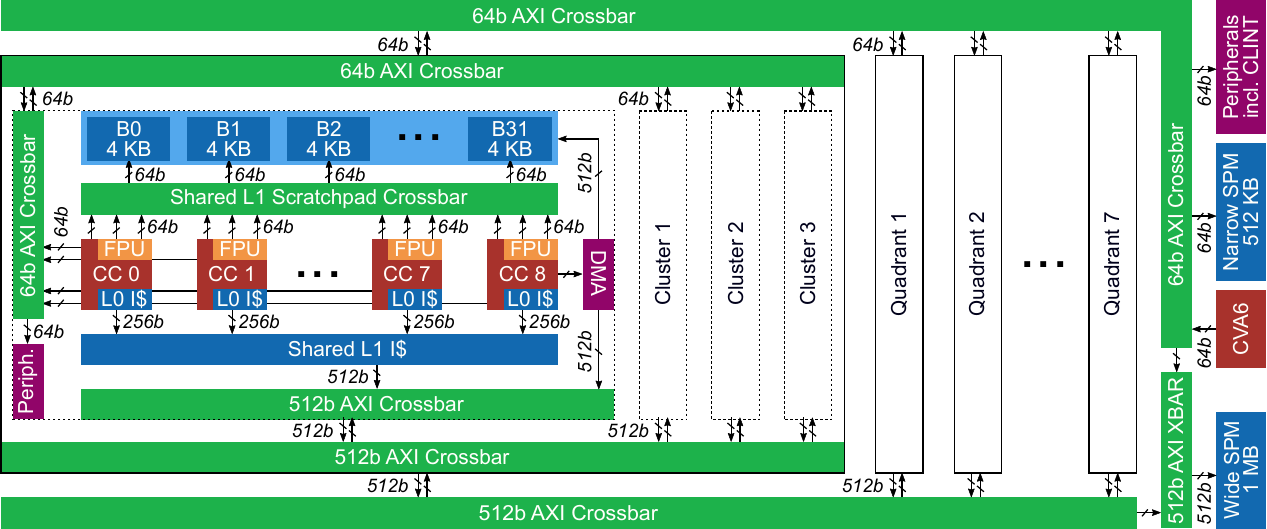}
\caption{Simplified block diagram of the Occamy \gls{soc}.}
\label{fig:occamy}
\end{figure*}

Concerning barrier synchronization, most architectures provide hardware barrier support among small clusters of cores, and rely on software-based synchronization between clusters. The A64FX CPU implements a hardware barrier per \textit{Core Memory Group} (CMG), which comprises of 13 cores \cite{a64fx}. Esperanto's ET-SoC-1 chip features fast local barriers within groups of 32 cores forming a \textit{Minion shire} \cite{ditzel2022}. Graphcore reports ``Hardware global synchronization in around 150 cycles on chip'', but implementation details are undisclosed \cite{knowles2021}. Occamy provides hardware barrier support within clusters. On top of this, we implement a job completion unit for fast synchronization across clusters, described in section \ref{sec:job_completion_unit}.

To the best of our knowledge, this section presented the SOTA in interrupt and barrier synchronization infrastructure: Occamy, detailed in the next section, represents a SOTA-aligned platform and baseline for our analysis.

\section{Background}
\label{sec:background}

\subsection{Occamy}

Occamy \cite{occamy} is an open-source chiplet-based architecture for data-parallel floating-point workloads. The system is highly configurable, such that the number of management cores, quadrants and clusters can be tuned at design time.

We employ a larger configuration w.r.t. that described in the paper, as represented in figure \ref{fig:occamy}.
It features a single CVA6 host core \cite{zaruba2019} coupled to a symmetric multi-processor accelerator comprising 288 energy-efficient Snitch\cite{zaruba2021snitch} cores. The accelerator cores are organized hierarchically in eight quadrants of four clusters each. Every cluster consists of eight compute cores with private double-precision \glspl{fpu} and one data mover core provided with a tightly-coupled \gls{dma} engine. All cores in a cluster have low-latency access into a 128 KB \gls{tcdm}, divided into 32 banks. Clusters and quadrants are interconnected by a 64-bit network for narrow \gls{lsu} accesses and a wide 512-bit network shared by the \gls{icache} and \gls{dma} subsystems.
At the top level, a 512 KB \gls{spm} and various peripherals are connected to the narrow network, and a 1 MB \gls{spm} is connected to the wide network. The wide and narrow networks are interconnected such that narrow requests can be forwarded onto the wide network.


\subsection{Fundamental offload tasks}
\label{sec:offload-tasks}

Assuming the host-centric heterogeneous model of execution, carrying out a job on the accelerator can be broken down into the following fundamental set of operations.

\textit{Communicate job information}. First of all, the host must inform the accelerator of the job to execute. The accelerator needs a pointer to the sequence of instructions representing the job, e.g. a function pointer. To reduce code duplication and size, a job may take arguments as input. For instance consider an AXPY job, whose arguments could be the length of the vectors to sum or the multiplicative constant $\alpha$. If the job takes any arguments, the host must communicate these to the accelerator as well.

\textit{Wakeup}. To harvest the benefits of heterogeneous computing, the accelerator should default to a low-power state until a job is received. Thus, part of the job offloading mechanism involves waking up the accelerator or a part of it. Interrupt-based wakeup mechanisms are preferred over polling-based mechanisms for their lower power consumption, by allowing to suppress any switching activity within the core's instruction pipeline.

\textit{Communicate job operands}. In addition to the job arguments, which were qualitatively described as to complete the specification of the job, the accelerator may consume other data as input. Let's once again take the AXPY job example. Pointers to the $\textbf{x}$ and $\textbf{y}$ vectors would typically be passed as arguments to the function, while the actual vector data would be loaded upon function execution. Both are inputs to the job and the distinction between these two types of inputs is fuzzy. Rather than a fundamental difference, practical considerations such as the size of the inputs lead to the separate treatment of \textit{arguments} and \textit{operands}.

\textit{Job execution}. Once job information and operands are available and the accelerator is awake, the actual computation can take place.

\textit{Communicate job results}. When the computation is over, any results which are needed for further processing must be transferred back to the host memory.

\textit{Notify job completion}. Finally, the host must be notified when the job is complete. The accelerator can then enter the low power state again, until the next offload.

\begin{figure}[t]
    \centering
    \includegraphics[width=\columnwidth]{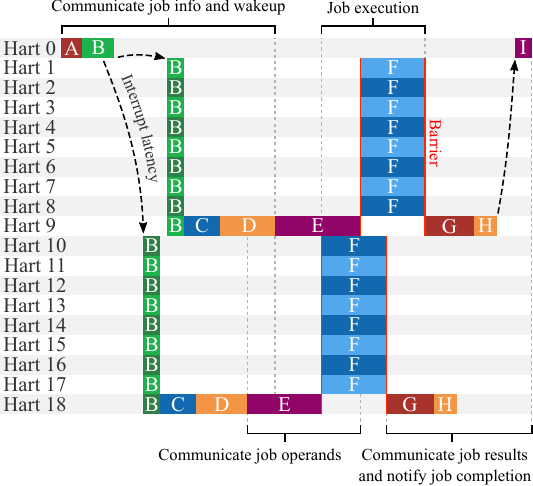}
    \caption{Example trace of the job phases in a 2-cluster Occamy system. Segments are drawn to qualitatively reflect actual observations. Dashed arrows represent interrupt latencies and red lines represent synchronization barriers. Brackets at the bottom and top of the figure group the phases which belong to the same fundamental offload task described in section \ref{sec:offload-tasks}.}
    \label{fig:offload-tasks}
\end{figure}

\section{Implementation}
\label{sec:implementation}

\subsection{Offloading in Occamy}
\label{sec:mpsoc-offloading-overheads}

We developed a bare-metal implementation of the host-centric heterogeneous execution model for Occamy. Hence, our baseline is not degraded by complex software layers. This is to present a lower-bound on the latencies and overheads involved in offloading as they arise in the hardware. Inefficiencies introduced by higher levels of software, e.g. operating systems, are not in the scope of this study and should be evaluated independently.

To decouple our analysis from CVA6's cache subsystem implementation
we make the following assumptions.
Instead of sending job operands from CVA6's \gls{dcache} to the accelerator TCDM, we assume these can be found in the on-chip \gls{spm} (e.g. as might be stored by a previous job). Analogously, instead of the job results' destination being CVA6's \gls{dcache} (for further processing on the host) we assume the destination to be the shared on-chip \gls{spm}. These assumptions are reasonable, and representative of well-optimized software, because the latency to move data between CVA6's cache and the accelerator's TCDM is similar to the latency between the TCDM and the shared \gls{spm}.

We will now go over the bare-metal implementation of the previously described offload process in Occamy. A representative example trace of this implementation is depicted in figure \ref{fig:offload-tasks}.

\textit{A) Send job information}. CVA6 writes job pointer and arguments at the base of cluster 0's TCDM memory.

\textit{B) Wakeup}. The low-power state of the accelerator can be controlled core-wise. Every core independently enters the low-power state by issuing a \gls{wfi} instruction. Wakeup is therefore achieved by sending an interrupt to every core. Upon receiving the interrupt, every core independently clears it.

\textit{C) Retrieve job pointer}. Once awake, every Snitch cluster, with the exception of cluster 0, will copy the job information stored in cluster 0's TCDM to its own TCDM. The \gls{dma} core initiates this operation using the tightly-coupled \gls{dma} engine. The number and size of the job arguments depends on the job, and thus so does the \gls{dma} transfer size. For this reason, every cluster first loads the job handler pointer from cluster 0 through a regular load. Once the DM core retrieves the pointer it invokes the job handler.

\textit{D) Retrieve job arguments}. The job handler statically carries the information on the size of the \gls{dma} transfer and programs the \gls{dma} accordingly. The job arguments' \gls{dma} transfer occurs in this phase.

\textit{E) Retrieve job operands}. Based on the job information retrieved in the previous step, every cluster independently initiates one or more \gls{dma} transfers to copy the job operands from the wide \gls{spm} into its TCDM. This is accomplished by the \gls{dma} core in the cluster. After initiating the transfers, the \gls{dma} core polls the tightly-coupled \gls{dma} engine to monitor the status of the transfers.

\textit{F) Job execution}. Once all job operands have been copied into the cluster's TCDM, the \gls{dma} core synchronizes with the compute cores through a hardware cluster barrier. The compute cores then proceed to execute the actual job computation.

\textit{G) Writeback job outputs}. Eventually, all compute cores in a cluster will be done with the job execution. At that point, they synchronize once again with the \gls{dma} core, which will initiate a \gls{dma} transfer to writeback the job outputs from its TCDM to the wide \gls{spm}.

\textit{H) Notify job completion}. Once all clusters have completed all the above steps, they must notify CVA6 of job completion. One core from every cluster, the \gls{dma} core, participates in a central-counter software barrier to synchronize all clusters. The central counter is stored in the TCDM of the first cluster. The last core to reach the barrier sends an interrupt to CVA6. As soon as each core has completed its part in the synchronization routine, it enters the low-power state by issuing a \gls{wfi} instruction.

\textit{I) Resume operation on host}. Once CVA6 receives the interrupt, it clears it and proceeds with the execution of the workload.

While this implementation is developed and presented for Occamy, it relies on platform characteristics which are present in most modern MPSoCs, that is: an integrated host processor, accelerator clusters (or cores) with local SRAM memories, and an on-chip network connecting both host and accelerator clusters.
As an example, both the ET-SoC-1 \cite{ditzel2022} and Wormhole \cite{ignjatovic2022} processors satisfy these requirements, replacing CVA6 respectively with the ET-Maxion and System Management cores, the Snitch clusters respectively with the Minion
Shires and Tensix cores, and both featuring a 2D mesh-based network connecting the host and accelerator clusters.
Thus, a similar offload framework to the one presented could readily be developed following our methodology, and the same optimizations presented in sections \ref{sec:multicast} and \ref{sec:job_completion_unit} could be applied to these platforms.

\subsection{Multicast-capable interconnect}
\label{sec:multicast}

\begin{figure}[t!]
    \centering
    \includegraphics[width=\columnwidth]{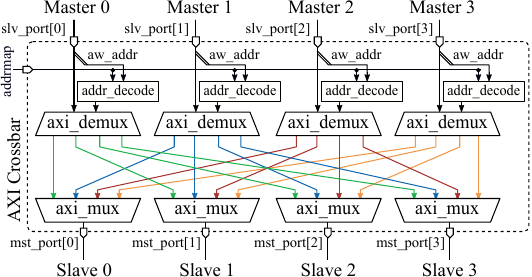}
    \caption{Block diagram of a 4x4 AXI XBAR.}
    \label{fig:xbar}
\end{figure}

By co-designing the hardware and the offload routines, we can improve the performance of heterogeneous applications. We demonstrate this by extending the Occamy interconnect with multicast capabilities.

Occamy's interconnect provides two paths for data movement: a narrow, 64-bit network for system configuration, synchronization and irregular data transfers, and a wide, 512-bit network for cache line and regular \gls{dma} transfers. In our configuration, each network is structured as a two-level tree of \glspl{xbar}. As can be seen in figure \ref{fig:occamy}, every four clusters are interconnected by two \glspl{xbar}, one narrow and one wide, to form a quadrant. Each of these \glspl{xbar} is connected to a higher level \gls{xbar}, interconnecting the quadrants and all other components in the system, such as the CVA6 core and system-level \glspl{spm}.

We extend the architecture of the baseline AXI \gls{xbar} described by Kurth et al. \cite{kurth2022} to enable multicast on Occamy's narrow interconnect. The architecture of the AXI \gls{xbar} is represented in figure \ref{fig:xbar}. At the interface, masters connect to the slave ports of the \gls{xbar}, and slaves to the master ports. When a write request arrives on a slave port, the destination address (\lstinline{aw_addr}) is compared with the address map of every master port (by the \lstinline{addr_decode} module) and the request forwarded to the slave connected on the matching port. On the other hand, a multicast write request carries multiple destination addresses, and multiple master ports may thus match the request.

\begin{figure}[t]
    \centering
    \includegraphics[width=\columnwidth]{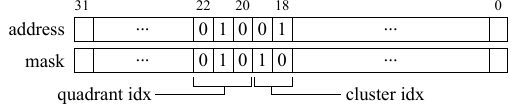}
    \caption{Encoding of multiple addresses.}
    \label{fig:multiaddr}
\end{figure}

To simplify the decoding logic, we exploit the fact that all clusters feature the same address space, offset by a constant \lstinline{0x40000} bytes from one cluster to another. We then adopt the following encoding to represent multiple destination addresses: the address field carries any one of the destination addresses; an additional signal is then added to the request, which serves as a mask on the address' bits. If a bit in the mask is set to 1, the corresponding bit in the address is interpreted as a don't care, i.e. it encodes both logic 0 and 1. Masking $n$ bits of the address allows to represent $2^n$ addresses. An example is provided in figure \ref{fig:multiaddr}. Here, bits \lstinline{[0, 17]} represent an offset inside the cluster's address space \lstinline{[0x0, 0x40000)}, bits \lstinline{[18, 19]} index one of the four clusters inside a quadrant, and \lstinline{[20, 22]} index one of the eight quadrants. In particular, the values in figure \ref{fig:multiaddr} address cluster 1 in quadrant 2. By masking bits 19 and 21, the multicast request encodes four addresses, each within a distinct cluster, and at the same offset within every cluster. Namely, the addressed clusters are at indices 1 and 3 in both quadrants 0 and 2.

The same representation can be used to encode the master ports' address maps. In fact, the mask encoding can be used to represent any interval whose length is a power-of-two and the start address is an integer multiple of the length, conditions which are satisfied in Occamy. The decoding logic then translates to the following condition:
\begin{lstlisting}[breaklines=true]
  match = &((req.mask | am.mask) | ~(req.addr ^ am.addr));
\end{lstlisting}

We extend the address decoder to incorporate this logic and the demux to simultaneously forward a request to multiple master ports. We synthesize the design using Synopsys' Fusion Compiler 2021.06 under
worst-case conditions at 0.72\,V and 125\,\textdegree C in GLOBALFOUNDRIES’ 12LP+ technology, with a target frequency of 1\,GHz. For an 8x8 \gls{xbar}, our extensions introduce an area overhead of 11\,kGE, less than 10\,\% of the original area, while still meeting the target operating frequency of 1 GHz.

We omit the details of a full design space exploration, as the multicast extension is not the central contribution of this work.
In this work, we focus on the end-to-end offload performance gains which can be attained when multicast support is provided. To this end, we co-design the \textit{Send job information}, \textit{Wakeup}, \textit{Retrieve job pointer} and \textit{Retrieve job arguments} phases with the multicast-capable interconnect.

As CVA6 can only send the job information to one cluster at a time, in the baseline implementation we chose to dispatch this information to a single cluster. We introduced the \textit{Retrieve job pointer} and \textit{Retrieve job arguments} phases to distribute this information to all other clusters in a second step. This is not the only possible implementation, however it improves on memory-level parallelism as CVA6's memory subsystem supports only a low number of outstanding write transactions.

With the multicast extension, we simultaneously distribute the job information to all clusters from CVA6's LSU and get rid of the \textit{Retrieve job pointer} and \textit{Retrieve job arguments} phases. The same goes for the \textit{Wakeup} phase. As the MCIP registers lie at the same offset in every cluster's memory map, we can write to them with a multicast transaction, simultaneously waking up all clusters.

We specifically target multicast as a broadcast extension would not be as widely applicable. As we will show in section \ref{sec:results}, not all workloads benefit from parallelizing onto all cores. In these cases, only a subset of clusters are employed, so broadcasting is not sufficient, and the flexibility of multicast is needed.

Finally, while our multicast implementation is designed for XBAR-based systems such as Occamy, many works in the literature cover multicast designs suited for a diverse range of \glspl{mpsoc} \cite{jerger2008, krishna2011, wang2009, ouyang2023}, including the ET-SoC-1 \cite{ditzel2022} and Wormhole \cite{ignjatovic2022} processors.
In all these cases, the mentioned offload phases can benefit from multicast support, as we will show in section \ref{sec:mcast-speedup}, although the exact improvements will depend on the interconnect latencies from the integrated host to each accelerator cluster, as well as the amount of parallel links and outstanding transactions supported by the interconnect, and their bandwidth.

\subsection{Job completion unit}
\label{sec:job_completion_unit}

\begin{figure}[t!]
    \centering
    \includegraphics[width=0.8\columnwidth]{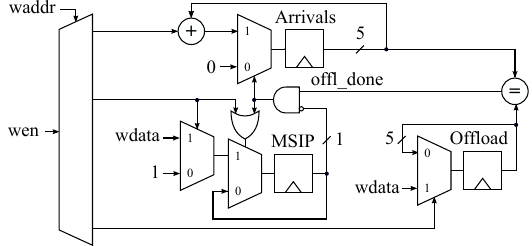}
    \caption{Job completion unit logic.}
    \label{fig:job_completion_unit}
\end{figure}

To improve the runtime and predictability of the global barrier synchronization in the \textit{Job completion notification} phase, we designed and integrated a dedicated job completion unit in the CLINT. The logic of the unit is represented in figure \ref{fig:job_completion_unit}. Upon an offload, CVA6 programs the \lstinline{offload} register with the number of clusters selected for offload. When a cluster completes a job, it writes to the \lstinline{arrivals} register which is atomically incremented as a side effect. When the \lstinline{arrivals} counter reaches the value set in the \lstinline{offload} register, the job is considered complete. If there is no pending software interrupt, the CLINT automatically fires one to notify CVA6, otherwise this will occur as soon as the previous pending interrupt is cleared. Finally, the \lstinline{arrivals} counter is automatically reset to zero for the next offload.

Intuitively, the logic involved is minimal and the \gls{ppa} overhead negligible w.r.t. the rest of the system.

Finally, multiple copies of this logic can be instantiated to support multiple outstanding jobs, e.g. to implement complex scheduling strategies such as task overlapping.
To this end, among the job information, CVA6 communicates a job ID, which is used to address the respective \lstinline{offload} and \lstinline{arrivals} registers, and is automatically set as the interrupt cause by the job completion unit for host inspection.

\section{Results}
\label{sec:results}

\subsection{Methodology}
\label{sec:methodology}

All experiments are conducted through cycle-accurate RTL simulations of the Occamy \gls{soc} using QuestaSim 2022.3. We instrument the source code with \lstinline{mcycle} CSR read instructions to mark the program segments we want to measure. In Snitch, these instructions execute in a single clock cycle, guaranteeing the method to be minimally intrusive. Core traces are logged to file during the simulation. We parse the trace files to extract the instrumented instructions together with their simulation timestamps. Through these timestamps we measure the runtime of the program segments of interest. The testbench drives the system clock at a frequency of 1 GHz, thus all runtimes reported in nanoseconds are in 1:1 correspondence with the runtime in CPU cycles.

We implemented six representative kernels for offload to the Snitch-based accelerator:
\begin{itemize}
    \item AXPY: a linear algebra level-1 workload from the BLAS library calculating $\alpha \cdot \textbf{x} + \textbf{y}$ on vectors of length $N$.
    \item Monte Carlo: a simple Monte Carlo integration approximating $\pi$ by randomly sampling $N$ points inside the unit square.
    \item Matmul: a linear algebra level-3 workload from the BLAS library calculating an $M \times N$ matrix $\textbf{C} = \textbf{A} \cdot \textbf{B}$, from matrices of size $M \times K$ and $K \times N$ respectively.
    \item ATAX: a linear algebra benchmark from the PolyBench suite, calculating $\textbf{A}^{t} \cdot \textbf{A} \cdot \textbf{x}$, between a matrix $\textbf{A}$ of size $M \times N$ and a vector $\textbf{x}$ of length $N$.
    \item Covariance: a data mining benchmark from the PolyBench suite, calculating the $M \times M$ covariance matrix from an $M \times N$ matrix.
    \item Breadth-First Search (BFS): a graph traversal algorithm included in the Graph500 suite, calculating the distance of all nodes in a graph from a selected node.
\end{itemize}
While these benchmarks are by no means exhaustive, they serve to demonstrate key findings that transcend a particular workload, as we will highlight in the following sections.

All workloads operate on double-precision floating-point operands and were compiled using a custom toolchain for Snitch based on LLVM 12.0.1 with \lstinline{-O3} optimization.

\subsection{Absolute impact on application performance}
\label{sec:absolute-overhead}

In this section, we evaluate the impact of the offload overheads on application performance.
As will be later shown, the presence of the offload phases indirectly affects the runtime of subsequent phases.
Thus, we cannot simply measure the offload overheads on the offloaded application in isolation.
Instead, to capture the full impact of offloading on application performance, we compare the runtime of an application executed directly on the device (ideal runtime), to the runtime of the same application offloaded from the host to the device (base runtime).
We estimate the offload overhead as the difference between the two runtimes.

\begin{figure}[!t]
    \centering
    \includegraphics[width=\columnwidth]{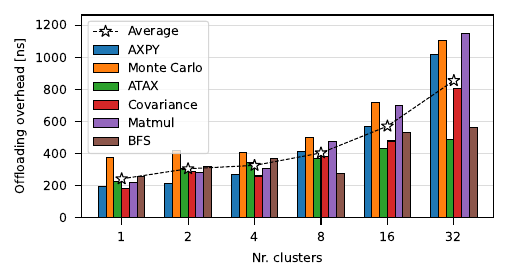}
    \caption{Offload overhead for several applications, when offloaded to a variable number of accelerator clusters.}
    \label{fig:offload-overhead}
\end{figure}

\begin{figure}[!t]
    \centering
    \includegraphics[width=\columnwidth]{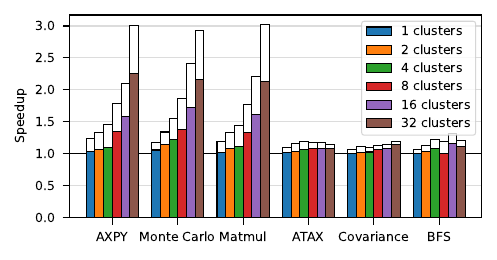}
    \caption{The white bars represent the ideal speedup attainable if the offload overheads could be completely eliminated. The colored fill levels indicate the speedup achieved with our extensions.}
    \label{fig:speedup-vs-nr-clusters}
\end{figure}

Figure \ref{fig:offload-overhead} reports the overhead measured on the selected applications, for various numbers of clusters.
On a single cluster, the average overhead is $\ResultOverheadSingleClusterMean$ cycles, with a standard deviation of $\ResultOverheadSingleClusterStddev$ cycles.
Independent of the application, the overhead consistently increases with the number of clusters, reaching a maximum of $\ResultOverheadMatmulMax$ cycles on a 32-cluster Matmul.

Interestingly, the difference between applications also increases, with a maximum standard deviation of $\ResultOverheadThirtyTwoClusterStddev$ cycles on 32 clusters.
This can be explained in light of the second-order effect offloading has on the runtime of non-offload phases.
As figure \ref{fig:offload-tasks} shows, the offload phases create an offset between the start times of Phase E on different clusters.
On the other hand, in the ideal scenario, Phase E would start simultaneously on all clusters, resulting in longer stalls as all clusters attempt to simultaneously access the shared interconnect and wide \gls{spm}.
Effectively, a part of the time spent on the offload phases is spared on contention stalls; up to as much time as the offset between Phase E on the first and last cluster.
In the case of the benchmarked Matmul kernel, the memory transfers and corresponding stalls are short, only a fraction of this maximum time.
The ATAX kernel, on the other hand, features larger transfers, and the entirety of the maximum time is saved, resulting in a lower effective overhead from offloading.
Clearly, the mentioned offset increases with the number of clusters, explaining the increasing difference between the Matmul and ATAX kernels in figure \ref{fig:offload-overhead}.

\subsection{Relative impact on application performance}
\label{sec:relative-overhead}

The white bars in figure \ref{fig:speedup-vs-nr-clusters} represent the speedup attainable if the offload overheads could be completely eliminated, for various applications and offload configurations.
This metric considers the offload overhead in relation to the offloaded application runtime. Offload overheads become negligible when compared to long application runtimes, peculiar to coarse-grained jobs, and will reflect in low ideal speedups.

We can immediately distinguish two groups of applications exhibiting a significantly different behaviour.
Speedups for the AXPY, Monte Carlo and Matmul kernels significantly increase with the number of clusters, in line with the trend observed on the offload overheads, reaching up to $\ResultIdealSpeedupMatmulMax \times$ on a 32-cluster Matmul.
On the other hand, the ATAX, Covariance and BFS kernels show near constant ideal speedups, topping out at $\ResultIdealSpeedupBFSMax \times$ on a 16-cluster BFS kernel.

The reason can be understood by examining figure \ref{fig:runtime}. The blue and green curves respectively show the base and ideal runtimes of the AXPY and ATAX jobs on the accelerator. After eliminating the offload overheads, the AXPY kernel responds to Amdahl's law: it can be sped up indefinitely by allocating more clusters to the job, albeit with diminishing returns. On the other hand, the runtime of the ATAX kernel still increases with the number of clusters, as the $\textbf{A}$ matrix and $\textbf{x}$ vector are broadcast sequentially to a higher number of clusters.
Thus, while the cycles associated to the offload overheads increase with the number of clusters, as seen in figure \ref{fig:offload-overhead}, they never represent a significant portion of the overall runtime of the ATAX kernel, resulting in low speedups.
The same is true for the Covariance and BFS kernels, which feature similar communication patterns.

What we conclude from our analysis is that, independent of the application, offloading incurs an overhead in the order of hundreds of cycles which can have a significant impact on fine-grained jobs. Furthermore, since this overhead increases with the number of clusters, we note that optimizing the offload overheads is most critical for large many-core accelerators.

\subsection{Restoring part of the ideally attainable speedup}
\label{sec:mcast-speedup}

In section \ref{sec:multicast}, we claimed that by co-designing the hardware and the offload routines we could reduce the offload overheads, thereby restoring a part of the ideally attainable speedups on the accelerator. To this end, we extended Occamy's interconnect with multicast support, designed the job completion unit and adapted the offload routines to take advantage of these extensions. To evaluate the impact of this effort, we compare the performance of the offloaded application with and without our extensions.

\begin{figure}[t]
    \centering
    \includegraphics[width=\columnwidth]{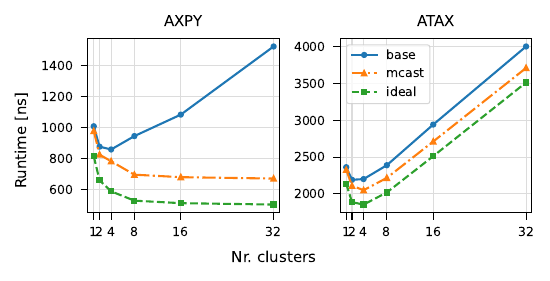}
    \caption{Base, ideal and improved runtimes of the AXPY and ATAX jobs on the accelerator for various numbers of clusters.}
    \label{fig:runtime}
\end{figure}

Figure \ref{fig:speedup-vs-nr-clusters} shows the actual speedup achieved through our extensions on several applications and for several offload configurations.
The same behaviour can be observed as was discussed for the ideal speedup in the previous section.
In fact, as can be seen from figure \ref{fig:runtime}, the runtime curves with our extensions (orange) closely track the ideal curves (green), offset only by a near-constant overhead centered at $\ResultMcastOverheadMean$ cycles, and featuring a small standard deviation of $\ResultMcastOverheadStddev$ cycles.
This overhead is associated with physical factors which cannot be trivially eliminated, such as the interrupt travel distance between host and accelerator.

Furthermore, the runtime of the AXPY application offloaded with our extensions does not feature a global minimum, as found in the baseline curve (blue).
This means that, thanks to our extensions, we are able to restore the ideal Amdahl-aligned behaviour discussed in the previous section, thanks to which we can always leverage additional clusters to improve the performance of the offloaded application.

We can quantitatively estimate how much of the ideally attainable performance we can restore thanks to our extensions.
For the AXPY, Monte Carlo and Matmul applications, we measure speedups within $\ResultIdealSpeedupFractionAXPYMonteCarloMatmulMin \%$ and $\ResultIdealSpeedupFractionAXPYMonteCarloMatmulMax \%$ of the ideally attainable speedups. For the ATAX, Covariance and BFS kernels, on the other hand, we measure fractions within the $\ResultIdealSpeedupFractionATAXCovarianceBFSMin \%$ and $\ResultIdealSpeedupFractionATAXCovarianceBFSMax \%$ range.

\begin{figure}[t]
    \centering
    \includegraphics[width=0.48\textwidth]{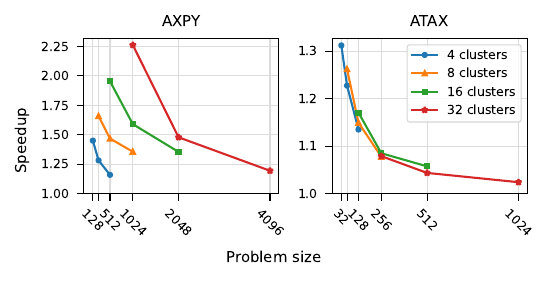}
    \caption{Speedup of the AXPY and ATAX jobs using our extensions over the baseline offload implementation. Various problem sizes are represented on the X-axis and different numbers of clusters are employed in the different curves.}
    \label{fig:speedup-vs-problem-size}
\end{figure}

In all of the previous experiments, we used a fixed problem size to highlight the dependency with the number of clusters (strong scaling), showing that optimizing the offload overheads is most beneficial for large many-core accelerators.
Figure \ref{fig:speedup-vs-problem-size} adds the dependency on the problem size to the picture. It displays the speedup with our extensions over the baseline for different problem sizes and number of clusters (or offload configurations). For every offload configuration, three problem sizes are chosen, such that the amount of work per cluster is the same in every configuration (weak scaling). 

In every configuration, we can observe that the speedup decreases as we increase the problem size. As expected, for larger workloads the overhead from offloading constitutes a smaller part of the overall runtime, confirming that fine-grained heterogeneous tasks benefit the most from our extensions.
The same effects observed on large workloads can also be expected for ``small'' jobs running inefficiently on the accelerator; the speedup reported in this section depends primarily on the relative magnitude of the job runtime on the accelerator and the offload overheads.
This may for instance be the case for heavily load-imbalanced workloads, where the overall accelerator runtime depends on the runtime of the slowest cluster.

These results are in line with this work's goal to enable new opportunities for extreme heterogeneous execution, allowing to take advantage of the heterogeneous capabilities of the hardware on a wider range of tasks.
In contrast to large coarse-grained jobs, such as deep neural network layers, which are already efficiently deployed onto existing heterogeneous platforms, our work enables fine-grained workloads to take advantage of heterogeneous hardware.

We can again observe different behaviours between the two classes of applications represented by the AXPY and ATAX kernels. For any fixed problem size, the speedup of the AXPY kernel with our extensions compared to the baseline increases as we offload to a larger number of clusters. This is the case also for the ATAX kernel at low problem sizes (points 64 and 128), but the trend gradually inverts as the problem size grows. At the 512 point, we observe a higher speedup in the 16 clusters configuration than the 32 clusters. We already observed in figure \ref{fig:speedup-vs-nr-clusters} that the speedup does not always increase with the number of clusters and explained this as being due to the fact that the job operands have to be broadcast to all clusters sequentially. We now see that this behaviour is exacerbated at larger problem sizes, as moving the data takes a larger part of the overall runtime, and the impact of the offload overheads becomes less significant on the total.

Finally, we observe a speedup greater than one in all experiments, showing that our extensions consistently enhance performance. They can thus be integrated directly in an offloading library and used seamlessly by the programmer, without compromising performance.

\subsection{Dissecting the overheads}
\label{sec:breakdown}

\begin{figure*}[t]
\centering
\includegraphics[width=\textwidth]{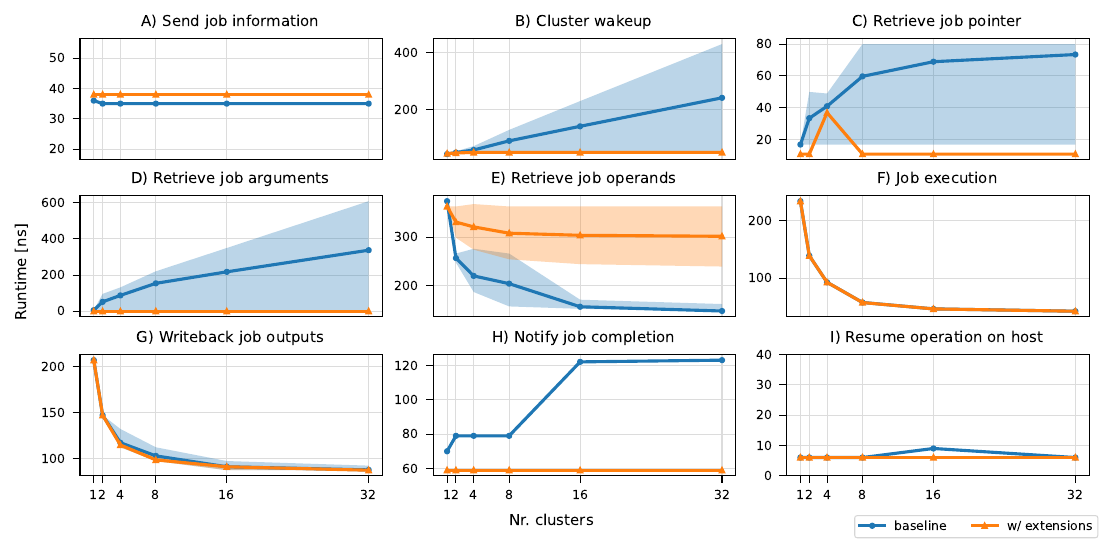}
\caption{Breakdown of the runtime of every phase of an AXPY job for various numbers of clusters with (orange) and without (blue) our extensions. The light blue and orange areas are delimited by the maximum (upper limit) and minimum (lower limit) runtimes across all clusters. The solid lines represent the average runtime across all clusters.}
\label{fig:breakdown}
\end{figure*}

We now attempt to provide a detailed quantitative breakdown of the overheads associated with offloading in Occamy. In section \ref{sec:speedup-model} we will develop a runtime model based on the breakdown presented in this section.

Figure \ref{fig:breakdown} breaks down the runtime of an offloaded AXPY application of size 1024 into the individual phases described in section \ref{sec:mpsoc-offloading-overheads}. Some of the phases are run by multiple processing elements and their runtime may differ across clusters. For such phases, we display the average, minimum and maximum runtimes over all clusters. The importance of the minimum and maximum statistics, over e.g. the standard deviation, will become clear in the course of the present and following sections.

\textit{A) Send job information}. This phase runs exclusively on CVA6. Its runtime is theoretically independent of the problem size and the number of clusters. The multicast and baseline implementations perform nearly the same, as the multicast extension only introduces two additional instructions, respectively to enable and disable multicasting.

\textit{B) Wakeup}. On the other hand, this phase shows a significant difference between the two implementations. The runtime is measured from the end of the previous phase to the moment each cluster wakes up after receiving an interrupt. The multicast extension allows CVA6 to send interrupts to all clusters simultaneously, whereas they are otherwise sent sequentially. There is thus barely any difference to wake up the first cluster, as shown by the minimum runtime, but the difference increases linearly as we wake up more clusters, as the maximum runtime confirms. Of the 47 cycles payed with multicast, 39 arise in the hardware as the write request exits CVA6's memory subsystem, propagates through the narrow network all the way to the clusters and wakes up each accelerator core.

\textit{C) Retrieve job pointer}. The baseline implementation requires every (remote) cluster to retrieve the job pointer from cluster 0. The minimum runtime, which coincides with the runtime on cluster 0, is expectedly (near) constant, while the maximum runtime increases with the number of clusters. The increase occurs only in two steps, when transitioning from one to two clusters and from one to two quadrants (four to eight clusters). The steps are due to the increased latency to access a remote cluster in the same quadrant, and a remote cluster in a different quadrant. Thanks to the offset created in phase B, the remote clusters do not have to contend access to cluster 0's TCDM, reason for which we do not see a linear increase in runtime. With the multicast extension, every cluster finds the job pointer in its own TCDM, where it was stored in phase A. Thus, the latency to retrieve the job pointer is that of a local TCDM access, matching the minimum runtime of the baseline implementation, and the overall runtime of the phase is (near) constant. 

\textit{D) Retrieve job arguments}. Presents a similar behaviour to the previous phase. The maximum runtime in the baseline implementation presents an intermediate behaviour between phase B) and C), owing to the fact that the offset created by the previous phase is not enough to fully avoid contention between clusters.

\textit{E) Retrieve job operands}. The total amount of data to move from the wide \gls{spm} to the TCDMs is the same (16 KiB) for every offload configuration. In the multicast implementation, all clusters start this phase at approximately the same time. As the wide \gls{spm} has a single read port, all clusters have to contend access to this resource, so the \gls{dma} transfers from every cluster will be granted sequentially. The maximum runtime of this phase, i.e. the runtime on the last cluster which is granted access to the \gls{spm}, will thus include the time to move all 16 KiB of data. The interconnect is designed such that multiple short \gls{dma} transfers perfectly interleave, thus taking the same amount of time as a single \gls{dma} transfer of combined length at the \gls{spm} interface, hence the maximum runtime stays constant, and can be modeled as:
\begin{equation}
t_{E}(N) = t_{setup} + t_{latency} + \frac{2 \cdot N \cdot 8}{bw}
\end{equation}
where $N$ indicates the AXPY vector length. Around 53 cycles are paid in instructions to setup the transfers of the \textbf{x} and \textbf{y} vectors, $t_{setup}$. An additional constant 55 cycles are ascribed to the round-trip latency of the \gls{dma} transfer, $t_{latency}$. This includes the time for:
\begin{enumerate}
\item the AR request to reach the \gls{spm}
\item the first R beat to return to the \gls{dma} engine
\item the \gls{dma} to forward the AW request and the first W beat to the TCDM
\item the B response to travel back to the \gls{dma} engine
\end{enumerate}
In addition to this latency, the \gls{dma} transfer takes one additional cycle per beat. The third addend calculates the number of beats required to transfer two double-precision vectors of length $N$ on the 512-bit wide network, featuring a bandwidth $bw$ of 64 bytes per cycle.

The baseline implementation shows a different maximum runtime behaviour. This is again due to the offset created in the previous phases, so no cluster is actually stalled waiting for the entirety of the data to be transferred. The overlap between the clusters' DMA transfers depends both on this offset and on the length of the DMA transfers themselves, i.e. on the problem size.
It is the dependency on the runtime of the previous phases and the second-order dependency on the problem size which make it more complicated to model the overall runtime of the offloaded application without our extensions. On the other hand, the data moved by each cluster is inversely proportional to the number of clusters employed. We would expect to see this behaviour reflected in the minimum runtime, assuming that the first cluster which is granted access to the \gls{spm} actually completes all \gls{dma} transfers (for both \textbf{x} and \textbf{y} vectors of the AXPY) in succession before any other cluster gets access. This is the case in the baseline implementation, thanks again to the offset created by the previous phases. In the multicast case, where all clusters initiate the first \gls{dma} transfer simultaneously, it so happens that no cluster sees both \gls{dma} transfers completing in succession without waiting on the \gls{dma} transfer of another cluster before the second \gls{dma} transfer is accepted. This is in line with the interconnect's design aimed at ensuring fairness.

\textit{F) Job execution}. The AXPY job execution phase shows a very predictable behaviour:
\begin{equation}
    t_F(n, N) = t_{init} + \frac{N}{throughput(n)}
\end{equation}
An upfront cost of 55 cycles is paid to configure and initialize the computation, $t_{init}$. It then takes 1.47 cycles on average to calculate each output vector element. These elements are evenly distributed to the eight compute cores in each of the $n$ clusters, yielding a throughput of $8 n / {1.47}$ elements per cycle.

\textit{G) Writeback job outputs}. The baseline implementation shows a similar behaviour to the \textit{Retrieve job operands} phase E. All clusters start the phase at differing times, due to the offset created in phase E itself, both in the baseline and multicast implementations. The minimum runtime matches in both implementations and coincides with the average runtime in the multicast implementation. That is, the initial offset between clusters is large enough to avoid any overlap in the \gls{dma} transfers of different clusters. Thus, differently from phase E, the runtime of this phase on every cluster corresponds to the runtime of a single \gls{dma} transfer:
\begin{equation}
    t_G(n, N) = t_{setup} + t_{latency} + \frac{N \cdot 8}{n \cdot bw}
\end{equation}
The first addend accounts for the cycles to set up the \gls{dma} transfer ($t_{setup}=21$), the second is the round-trip latency of the transfer, as seen also in phase E, and the last term adds one cycle for every beat required to send the $N/n$ double-precision elements calculated by each cluster on the 512-bit bus.
The deviation between the maximum and average runtimes of the baseline implementation is due to the overlap between phase E and phase G on different clusters, which occur when the offset between clusters is large. A cluster involved in phase G may observe a stall caused by the \gls{dma} transfer of a cluster involved in phase E. This is another instance of the runtime of two phases coupling in a non-trivial way.

\textit{H) Notify job completion}. In the baseline implementation, clusters are woken up from the highest index to the lowest, such that cluster 0, containing the barrier counter, arrives on the barrier last. This way we can overlap the longer latencies of the remote clusters in incrementing the counter with the offsets between clusters. The runtime of this phase, measured from the last core starting this routine (cluster 0) to CVA6 waking up, is nearly constant up to 16 clusters. Beyond 16 clusters, the offset between clusters shortens to the point that it is no longer large enough to hide the latency of remote clusters incrementing the central counter, resulting in an increase of the measured runtime. The multicast implementation would suffer even more of this issue, as it is not possible to enforce the arrival order of clusters on the barrier. We thus employ the job completion unit to achieve a predictable, constant runtime for this phase. By eliminating the need for atomic transactions, the runtime is also slightly improved.

\textit{I) Resume operation on host}. Exclusively run on CVA6, it is independent of both number of clusters and problem size.

While the exact numbers presented in this and the following sections are dependent on the target evaluation platform, as mentioned in section \ref{sec:mpsoc-offloading-overheads} the same offload procedure can be ported to different \glspl{mpsoc} and a similar analysis could thus be developed following our methodology.
By further breaking down and analyzing the offload process in its individual phases, exposing the connection between the runtime of each phase and the underlying hardware, and presenting the modalities in which the different phases ``compose'' to determine the overall runtime of an offloaded job, elaborated in section \ref{sec:speedup-model}, we believe that our work can serve as a methodological reference for estimations in other \glspl{mpsoc}.

\subsection{Offload runtime model}
\label{sec:speedup-model}

\begin{figure}[t]
    \centering
    \includegraphics[width=\columnwidth]{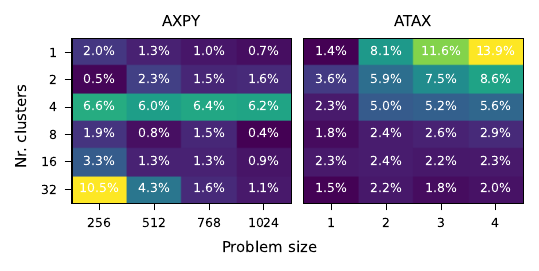}
    \caption{Relative error of the offloaded application runtime models for various problem sizes and numbers of clusters, calculated as $|t-\hat{t}|/t$. The problem size stands for the $N$ parameter in the case of AXPY and for the $M$ parameter in the case of ATAX.}
    \label{fig:model_accuracy}
\end{figure}

In the previous section we gave an intuition on the difficulties involved in modeling the runtime of applications offloaded with the baseline routines. A side benefit of the multicast implementation is that it simplifies this modeling. For the mentioned reasons, we develop the following analysis considering the improved offload implementation only.

As we can observe from figure \ref{fig:breakdown}, in the multicast implementation, the average, min and max curves (approximately) coincide for all but the \textit{Retrieve job operands} phase, i.e. the runtime is nearly the same on every cluster. This allows us to model the runtime of an AXPY offload of size $N$ onto $n$ clusters as:
\begin{equation}
\label{eq:abstract-model}
    \hat{t}(n) = \sum_{p\in\interval{A}{I}} \max_{i\in\interval[open right]{0}{n}} t_{p}(n, N, i)
\end{equation}
where the individual phase runtimes $t_p$ are a function of the number of clusters selected for offload, the problem size and the particular cluster measured.

Plugging in the numbers from section \ref{sec:breakdown} into equation \ref{eq:abstract-model}, we obtain the following quantitative model:
\begin{equation}
    \hat{t}(n) = 400 + \frac{N}{4} + \frac{2.47 \cdot N}{n \cdot 8}
\end{equation}
where e.g. 400 results from the sum of all constant phases (A, B, C, D, H, I) and the constant components of phases E, F and G.
According to this model, the offloaded AXPY kernel responds to Amdahl's law. The speedup attained by scaling the computation to multiple accelerator clusters is limited by the serial fraction, made up also by the offload overheads.

By developing a similar breakdown analysis to that presented in section \ref{sec:breakdown} for the ATAX kernel, we can derive a model for the runtime of an offloaded ATAX application of size $N \times M$, following the same methodology used for the AXPY kernel:
\begin{equation}
\label{eq:atax-model}
    \hat{t}(n) = 566 + 3.98 \cdot N \cdot M + \frac{2.9 \cdot N}{n \cdot 8} + \frac{N \cdot (1+M)}{8} \cdot n \\
\end{equation}
This application belongs to the second class identified in section \ref{sec:relative-overhead}: it does not follow Amdahl's law directly. Instead, in addition to a sequential fraction, independent on the number of clusters, and a parallel fraction, inversely dependent on the number of clusters, equation \ref{eq:atax-model} also presents a term which depends linearly on the number of clusters. This term, dominant for large numbers of clusters, explains the asymptotic behaviour observed in figure \ref{fig:runtime}.

We validate the accuracy of our models on a variety of problem sizes. Validation results are shown in figure \ref{fig:model_accuracy}.
The error is consistently lower than $\ResultMaxModelingErrorRounded \%$, proving that our models can be used to accurately estimate offloaded application runtimes.

While these models cannot be directly applied to other kernels, the methodology and analysis we presented can be readily used to derive offload runtime models for different kernels.
In particular, we note that the runtime of the offload phases A, B, C, D, H and I is mostly independent of the offloaded job, while the runtime of phases E and G can be calculated by adapting the formulas in section \ref{sec:breakdown} to the target application.
On the other hand, models \cite{williams2009, li1988, sun1990, yavits2014} exist which can be used to estimate the runtime of phase F.
These techniques can be used in conjunction with the model of the offload phases proposed in this work to derive complete and accurate models of offloaded application runtime.


\section{Conclusion}
\label{sec:conclusion}

In conclusion, we developed a fully open-source heavily-optimized bare-metal offloading software framework on a state-of-the-art open-source \gls{mpsoc}.

We used this framework to evaluate the impact of the offload overheads on the speedup of a real application, demonstrating the importance of reducing these overheads and making informed offload decisions.

We optimized the offload routines by extending the hardware with a multicast-capable interconnect and a job completion unit. By means of these optimizations we measured up to $\ResultMaxSpeedup \times$ speedups on offloaded applications, and were able to restore a significant fraction of the ideally attainable speedups, greater than $\ResultIdealSpeedupFractionAXPYMonteCarloMatmulMin \%$ in all our experiments.

Finally, we presented a detailed cycle-accurate breakdown of the offload overheads in a heterogeneous \gls{mpsoc}, setting a transparent baseline for future comparisons. We used these measurements to develop highly-accurate analytical models of offloaded application runtime, with relative errors consistently below $\ResultMaxModelingErrorRounded \%$ for small problem sizes. Such models could be used to formulate the offload decision as an optimization problem and analytically derive optimal offload parameters.

\ifCLASSOPTIONcompsoc
  \section*{Acknowledgments}
\else
  \section*{Acknowledgment}
\fi

This work has been supported in part by ‘The European Pilot’ project under grant agreement No 101034126 that receives funding from EuroHPC-JU as part of the EU Horizon 2020 research and innovation programme.


\bibliographystyle{IEEEtranS}
\bibliography{refs}


\vspace{-3em}

\begin{IEEEbiography}[{\includegraphics[width=1in,height=1.25in,clip,keepaspectratio]{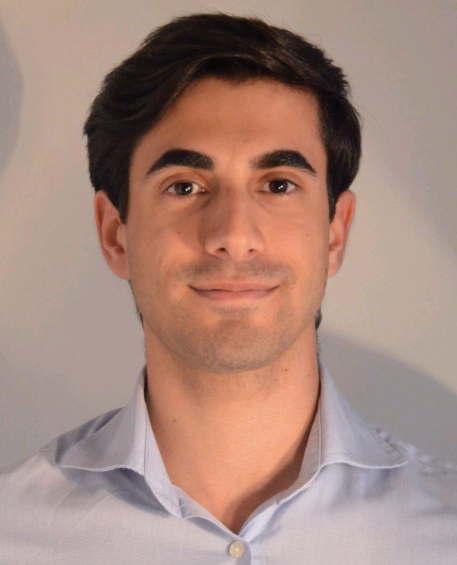}}]{Luca Colagrande}
received his BSc degree from Politecnico di Milano in 2018 and his MSc degree from ETH Zurich in 2020.
He is currently pursuing a PhD in the Digital Circuits and Systems group of Prof.
Benini.
His research focuses on the co-design of energy-efficient general-purpose manycore accelerators for machine learning and high-performance computing applications.
\end{IEEEbiography}

\vspace{-2em}

\begin{IEEEbiography}[{\includegraphics[width=1in,height=1.25in,clip,keepaspectratio]{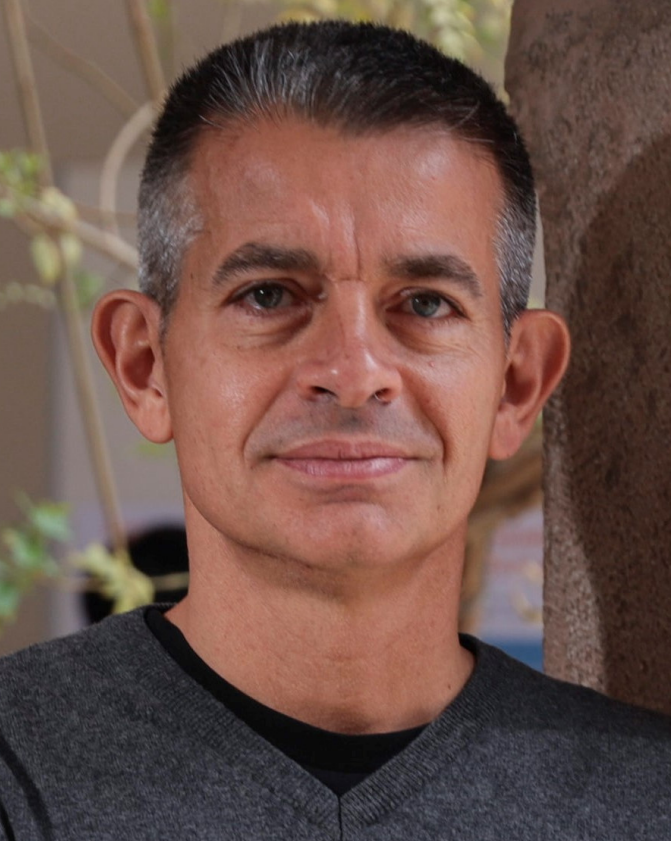}}]{Luca Benini}
holds the chair of digital Circuits and systems at ETHZ and is Full Professor at
the Universita di Bologna.
He received a PhD from Stanford University.
Dr. Benini’s research interests are in energy-efficient parallel computing systems, smart sensing micro-systems and machine learning hardware.
He is a Fellow of the ACM and a member of the Academia Europaea.
\end{IEEEbiography}
  
\end{document}